# GRACE – gravity data for understanding the deep Earth's interior

**Mioara Mandea [1,*], Véronique Dehant [2] and Anny Cazenave [3]**

[1] CNES – Centre National d'Etudes Spatiales, Paris, France; mioara.mandea@cnes.fr
[2] Royal Observatory of Belgium, Brussels, Belgium; veronique.dehant@oma.be
[3] LEGOS – Laboratoire d'Études en Géophysique et Océanographie spatiales, Observatoire Midi-Pyrénées, Toulouse, France: anny.cazenave@legos.obs-mip.fr

**\*** Correspondence: mioara.mandea@cnes.fr; Tel.: +33 6 76 28 17 85

20 December 2020

**Abstract:** While the main causes of the temporal gravity variations observed by the GRACE space mission result from water mass redistributions occurring at the surface of the Earth in response to climatic and anthropogenic forcings (e.g., changes in land hydrology, in ocean mass, in mass of glaciers and ice sheets), solid Earth's mass redistributions are also recorded by these observations. This is the case, in particular, for the Glacial Isostatic Adjustment (GIA) or the viscous response of the mantle to the last deglaciation. However, it is only recently showed that the gravity data also contain the signature of flows inside the outer core and their effects on the core-mantle boundary (CMB) [1-2]. Detecting deep Earth's processes in GRACE observations offers an exciting opportunity to provide additional insight on the dynamics of the core-mantle interface. Here, we present one aspect of the GRACEFUL (GRavimetry, mAgnetism and CorE Flow) project[1], i.e. the possibility to use the gravity field data for understanding the dynamic processes inside the fluid core and core-mantle boundary of the Earth, beside that offered by the geomagnetic field variations.

**Keywords:** GRACE satellite; gravity field; magnetic field; core-mantle boundary;

## 1. Introduction

Understanding the flow in the Earth's liquid outer core and its interactions with the lower mantle, as well as its effects on global Earth's observables such as the Length-of-Day (LoD) as well as gravity and magnetic fields is an important topic. In order to estimate core flows, a knowledge of the core magnetic field and its temporal variations at the core-mantle boundary is needed together with specific hypothesis about the flows. Core flows is a fundamental property of Earth's internal dynamics, but remains relatively poorly constrained, in particular its impact on the core-mantle boundary (CMB). To address this topic, a multidisciplinary approach is needed. In a recently granted project by the European Research Council (ERC), called GRACEFUL (GRavimetry, mAgnetism and CorE Flow), we propose an integrative approach based on observations of the Earth's magnetic and gravity fields, and LoD. In this special issue dedicated to GRACE (Gravity Recovery and Climate Experiment) missions, we focus mainly on how different factors that cause temporal variations of the gravity field (including atmospheric loading, land hydrology, land ice loss and oceans mass change) can be estimated in order to obtain a residual signal that can be interpreted as the deep Earth's signature and that can be correlated with geomagnetic field variations. The approach is mainly data–driven, mostly provided by space missions.

---

[1] https://cordis.europa.eu/project/id/855677/en

With the advent of the space era about 50 years ago, determining the orbit of artificial satellites has become an imperative. Satellite orbits deform in a complex way over a broad range of time scales, mostly in response to the Earth's gravity field. Celestial Mechanics theory together with "tracking" measurements between the ground geodetic stations and satellites have been used to determine the satellite orbits from which models of the Earth's gravity have been derived. This has revealed that the gravity field is much more complex than previously thought, reflecting the heterogeneous distribution of the matter inside, at the surface and above the surface of the planet. Since many decades however, only the "static" (i.e., time invariable) part gravity field has been measurable, except for the temporal change of the very low degree harmonics of the spherical harmonics expansion of the field [3-5]. The situation drastically improved in 2002 with the launch of the GRACE mission [6-7]. For the first time, GRACE provided gravity field solutions with global coverage, allowing to measure temporal changes of the Earth's gravity with unprecedented spatio-temporal resolution and precision.

The "static" Earth's gravity field represents the present heterogeneous distribution of the matter inside the Earth and on its surface. One of the gravity field equipotential surfaces forms the geoid, an equipotential surface that approximates the mean sea level at rest. Since the early 1970s, observations provided by artificial satellites have revealed that the geoid shape is much less regular than previously thought. Interpretation of the long wavelength undulations combined with seismic measurements of internal density variations has led to major advances in our understanding of the large-scale convective structure of the Earth's mantle [e.g., 8]. The next major step came in the early 2000s, with the ability to detect temporal changes of the gravity field, a challenge since the static gravity field represents 99% of the total gravity signal. With a temporal resolution of ~ 1 month and a spatial resolution of ~ 300 km, GRACE-based gravity variations mostly reflect spatio-temporal variations of mass distributions occurring among the surface fluid envelopes of the Earth (atmosphere, oceans, terrestrial waters and cryosphere) in response to climate change and variability (both from natural and anthropogenic sources) and to direct human interventions [7]. New data continue to be added to the GRACE record since the launch of the GRACE Follow-On (FO) mission in 2018.

GRACE and GRACE FO also measure solid Earth's processes such as co-seismic and post-seismic crustal deformations [e.g., 9-10], as well as large-scale readjustment of the Earth's mantle to the last deglaciation (the so-called Glacial Isostatic Adjustment - GIA) [e.g., 11]. GRACE has also the capability of detecting lithospheric deformations and associated gravity changes due to on-going climate-related land ice melt [12-13]. While hundreds of scientific articles have been published during the past few years on the GRACE-based superficial mass redistributions, mass redistributions occurring deeper in the Earth have been the object of only a few investigations [1-2]. These studies also motivated the GRACEFUL project. Here, we propose to revisit the possibility to detect deep Earth's signals in the GRACE data using updated data sets that those in [1]. This paper needs to be seen as a short contribution to enlighten the possibility of using space gravity data to gain better knowledge in core dynamics. We first briefly discuss magnetic field observations (Section 1) and show that the deep interior contribution in gravity field (Section 2) may be correlated with the geomagnetic field (Section 3). In the last section (Section 4), we discuss future prospects.

**2. Materials and Methods**

*2.1. Magnetic field data analysis*

The study of the Earth's deep interior is in a period of significant evolution. The accumulation of seismic data provides important information on the material and physical properties of the core, however, only indirect observations are available on the dynamics of the Earth's fluid iron-rich outer core. Apart from seismic waves, the magnetic and gravity fields, as well as the Earth's rotation provide invaluable information on processes occurring in the Earth's deep interior.

The geomagnetic field measured above the Earth's surface by different satellites or by magnetic observatories, results from sources that are both internal and external to the planet. Dynamo action in the liquid outer core and crustal magnetization in the upper layers of the Earth constitute the internal sources, whereas electrical current systems in the ionosphere and magnetosphere belong to the external sources. The external ionospheric current systems are very complex in nature and appearance, and varies rapidly with location and time. In the region in which satellite measurements are usually taken (CHAMP[2] and Swarm[3] satellites fly between altitudes of roughly 300-500km), one important layer is the E-region, where electric currents flow on the dayside, in the altitude range of 90–150 km. Unless applying a dedicated processing, the contributions from this region would be interpreted as internal sources by a satellite. It is then crucial to have an understanding of the spatio-temporal scales associated with the contributions to the magnetic field from all the different sources.

The best possible models for the non-core sources must be used in order to extract the core field signal from satellite magnetic field measurements. Traditionally, the magnetic field is assumed to be measured in a source free region, thereby allowing for a description in terms of a potential field. This potential is typically expressed in terms of a spherical harmonics expansion. Two recent magnetic field models, GRIMM [14] and CHAOS [15], are worth to consider. However, as shown in Figure 1, significant differences at the level of 10 nT between both fields can be seen in the polar areas, indicating an insufficient parametrisation of the external field contributions in these regions. The leakage of external magnetic field signatures into the internal magnetic field model coefficients produce large-scale patterns. The cause for this leakage may be due to different data selection applied in the model derivations.

---

[2] https://isdc.gfz-potsdam.de/champ-isdc/
[3] https://earth.esa.int/web/guest/swarm/data-access

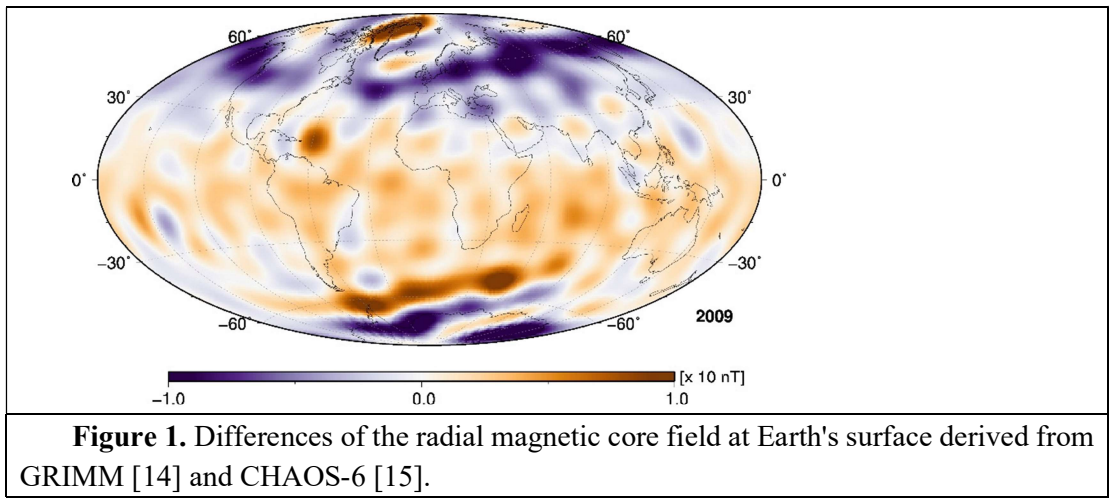

**Figure 1.** Differences of the radial magnetic core field at Earth's surface derived from GRIMM [14] and CHAOS-6 [15].

To choose a magnetic field model or to compute such a model using magnetic observatories and satellites data is a very challenging task. It is imperative to focus on a rigorous separation of external and internal geomagnetic field sources and their individual temporal characteristics, in particular from intra-annual to inter-annual periods, as well as their noises in the same way as [16-17]. In the following, we consider the CHAOS-6 model, which describes the core magnetic field and its temporal variations, the secular variation and secular acceleration [15].

*2.2. Gravity field data analysis*

We analysed monthly time series of GRACE gravity fields, over the period January 2003 to December 2015, and extracted the deep solid Earth's signal supposed to be correlated to the core magnetic field, after correcting for well-known other processes causing regional gravity variations. The latter include atmospheric loading, hydrological changes in river basins on land, glaciers and ice sheet mass changes, ocean mass changes due to land ice melt and ocean circulation changes, and GIA [18].

To extract the time-variable deep Earth's signal from the GRACE data, we used the most-up-to-date knowledge about each contribution (from the atmosphere, oceans, ice sheets and glaciers, terrestrial waters...) and removed the corresponding signal from the GRACE fields over their geographical areas of interest. We detail below the various data sets that we used in what follows, as well as the methodology that we have adopted to remove the various components.

2.2.1. GRACE fields

We use the ensemble mean of GRACE gravity field dataset processed by [19][4], available publicly[5]. This dataset consists of an average of five different GRACE solutions. They are based on spherical harmonics solutions provided by five different processing centers: CSR (Center for Space Research at Texas University), GFZ (GeoForschungsZentrum), JPL (Jet

---

[4] ftp.legos.obs-mip.fr/pub/soa/gravimetrie/grace_legos

[5] ftp.legos.obsmip.fr/pub/soa/gravimetrie/grace_legos

Propulsion Laboratory), TUG (Technical University of Graz), and GRGS (Groupe de Recherche de Géodésie Spatiale). The "Release 5" solutions are used for the CSR, GFZ and JPL data sets. "Release ITSG 2016" and "Release 3.3" are used for the TUG and GRGS solutions, respectively. For all GRACE solutions, the spherical harmonic coefficients up to degree 60 were used for the conversion to gridded mass anomalies. The post-processing applied by [19] includes: (i) the addition of independent estimates of the geocenter motion and Earth's oblateness as these quantities are either not or poorly observed by GRACE, (ii) a filtering for correlated errors characterized by north-south stripes, and (iii) a correction for the GIA. Different solutions have been considered in [19] for the geocenter, $C_{20}$ and GIA corrections in order to assess the GRACE data uncertainties. Here, we consider the gridded version of the ensemble mean of the five available solutions with the geocenter motion correction [20], $C_{20}$ correction [21] and GIA model [22]. The GRACE data cover the time span January 2003-December 2015.

### 2.2.2. Atmospheric loading and associated ocean bottom pressure

Atmospheric loading and associated ocean bottom pressure effects on GRACE solutions are accounted for by the processing centers, using a priori background ocean and atmospheric models. However, different model corrections are used in the GRACE solutions considered here (hence the ensemble mean). To cope with this, in [19] these models are added back to each GRACE solution, so that one can use our own ocean and atmosphere background models and further removed the same model for all solutions (see details in [19]). Atmospheric loading is removed in all available GRACE solutions, using atmospheric pressure fields from reanalyses of the European Center for Meteorological Weather Forecast (ECMWF).

### 2.2.3. Hydrological component

The signature of terrestrial waters is the dominant signal in GRACE data over land, in addition to ice mass change of glaciers. Terrestrial waters include surface waters (rivers, lakes, wetlands and man-made reservoirs), canopy, soil moisture, underground waters, and snow pack. Mass variations of these components are mainly driven by climate change and variability as well as human activities such as water extraction in aquifers for crop irrigation and domestic use, and reservoir construction on rivers. Multi-decadal trends of continental water mass are dominated by decreases due to groundwater depletion and by increases due to the construction of new reservoirs. Inter-annual variations are mainly driven by natural climate variability. Associated terrestrial water storage changes are routinely estimated from GRACE [e.g., 23-25]. However, to remove surface mass variations from GRACE data for further extracting the deep Earth's signal, independent data are needed. We used the most-up-to-date global hydrological models, in particular the latest version of the Water Gap Hydrological Model [26] that models natural variability in addition to human intervention.

### 2.2.4. Land ice component

Land ice melt associated with current global warming is an important contribution to the GRACE fields. Global glaciers are currently melting [27] and the ice sheets (Greenland and Antarctica) are losing mass at an accelerated rate [e.g., 28 and also the IMBIE 2 project[6]]. Three main methods are used to estimate the mass balance of the ice sheets: (1) measurement of changes in elevation of the ice surface over time from radar altimetry, (2) the mass budget or Input-Output Method, which involves estimating the difference between the surface mass balance and ice discharge, and (3) the redistribution of mass using GRACE. For the glaciers, GRACE is also used but the main methodologies to estimate mass change are in situ measurements combined with models. For both the glaciers and ice sheets, we used estimates based on non-GRACE approaches as described in detail in the WCRP Global Sea Level Budget Group [29].

2.2.5. Ocean mass change

The dominant signal for ocean mass change is the one due to the present day land ice melt. It is routinely measured by GRACE [e.g., 29]. The best alternative to obtain independent data is the use of gridded sea level fields from satellite altimetry corrected for ocean thermal expansion. Six different altimetry data sets are available, including the most recently improved one in the context of the ESA Climate Change Initiative project [30]. Several ocean thermal expansion data sets are also available from XBT (Expendable Bathythermograph) and Argo (international program) profiling float measurements (temperature, salinity, currents, and bio-optical properties) [31]. These data sets (altimetry and thermal expansion) are described in detail in [29].

2.2.6. GIA and other static factors

GIA induces important regional trends on GRACE solutions (e.g., [11-12]). To correct GRACE data for the GIA, global models are used. These models depend on the deglaciation history and mantle viscosity profile. In [19] three different solutions are considered to estimate model discrepancies and their impact on GRACE solutions. The ensemble mean GRACE solution considered here is corrected for the ICE6G-C GIA model [32].

2.2.7. Remaining effects – core effects

The gridded GRACE-based gravity time series were corrected for all effects described above (land water storage, land ice, and GIA). We call residual gravity field the corrected GRACE-based gridded time series. We have applied Empirical Orthogonal Function (EOF) analysis decomposition [33] to the residual GRACE gravity field to investigate the signal and its spatio-temporal variability that remains after applying all above corrections. This approach allows identifying the different modes of spatial and temporal variability of a signal. The first modes are associated with the largest variance of the total signal and in principle represent their dominant spatio-temporal components. The dominant inter-annual gravity signal (first EOF mode) is the one that correlates with the residual magnetic signal.

The gravity residuals include the signature of processes occurring in the core and at the core-mantle boundary (CMB). In addition to the topography generated by the convection in the mantle [e.g., 34], the CMB may be in a dynamic equilibrium state, mainly controlled by dissolution - corrodation - penetration - infiltration of the liquid core alloy in the overlying mantle silicate rocks [2], as well as a compaction process induced by the core flow on the mantle [35], as well as by a compaction process induced by the core flow on the mantle (also explained in [34]).

---

[6] http://www.climate-cryosphere.org/media-gallery/1626-shepherd

## 3. Results

In a pioneering article, [1] reported the existence of a correlation between spatio-temporal variations of the magnetic and gravity fields at inter-annual timescale. This correlation was later interpreted as reflecting processes occurring at the CMB [2]. The measured anomalies of a few tens $nT/y^2$ in the core magnetic field secular acceleration and of hundreds of nGal in the gravity field, obtained from the highest-sensitivity satellite measurements, are compatible with the estimated effect of dissolution-infiltration process.

We have revisited the results published by [1], mainly to investigate how the two fields correlate for a longer period of time. In [1], the considered time span starts in August 2002 and ends in August 2010. Here, the considered period is from January 2003 to December 2015. In addition, the models used for the corrections of the observed fields are different from the previous study. For the magnetic field, we have used the series of field variations as provided by CHAOS-6 model[7], providing the most precise values of the secular acceleration of the vertical magnetic field component. For the gravity field, the series are obtained from the pre-processed LEGOS V0.94 mean value model[8]. Note that, for the few missing months, a cubic spline is applied with respect of the annual, semi-annual cycles and the trend, and to minimize the sub-annuals signals, a Hamming window of 15 months is applied. Over the indicated period, both models are truncated at degree / order 8, in order to characterize the large spatial scale. These two models are considered for reconstructing the corresponding magnetic and gravity time series on a global grid defined by 10° in geographic latitude and 20° in geographic longitude; this can be considered to be a well-distributed network of "virtual magnetic and gravity observatories" (VMGOs) located at the center of each cell. At the position of each observatory, series of both fields are computed as monthly means.

---

[7] http://www.spacecenter.dk/files/magnetic-models/CHAOS-6/
[8] ftp.legos.obs-mip.fr/pub/soa/gravimetry/grace_legos

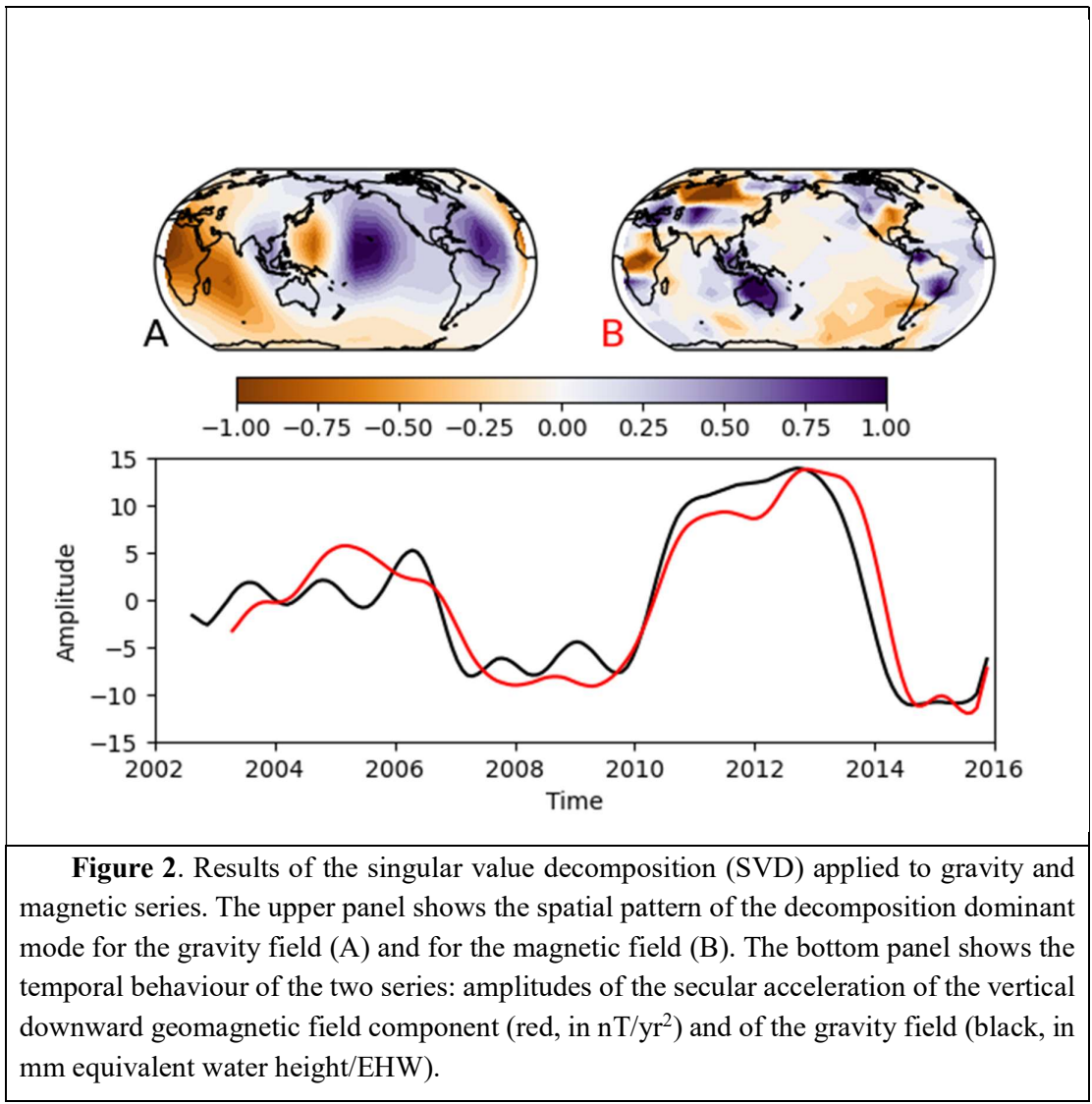

**Figure 2**. Results of the singular value decomposition (SVD) applied to gravity and magnetic series. The upper panel shows the spatial pattern of the decomposition dominant mode for the gravity field (A) and for the magnetic field (B). The bottom panel shows the temporal behaviour of the two series: amplitudes of the secular acceleration of the vertical downward geomagnetic field component (red, in nT/yr$^2$) and of the gravity field (black, in mm equivalent water height/EHW).

Considering that the core processes are large-scale phenomena, the common variability of the magnetic and gravity fields is investigated at global scale. We choose to apply, as in [1], the singular value decomposition (SVD) technique, which is generally used to retrieve common variability modes from two data sets [36]. This can be seen as a generalization of the Empirical Orthogonal Function decomposition, which isolates common variability modes in a set of time series. Applying this method allows us to retrieve this common variability for all the VMGOs, by decomposing the two sets of time series into a number of modes of common variation. Each mode consists of a spatial pattern and a time series for each dataset. The temporal variations are smoothed with a 6-month window. Figure 2 shows temporal and spatial variations of the part of the gravity and magnetic fields interpreted as originating in the core after minimizing the contributions from surface and external sources, and subtracting the annual signal and the mean. It is possible to detect large-scale magnetic and mass distribution fluctuations. The associated time series show a slow oscillation at the sub-decadal time-scale, consistent with the suddenness of geomagnetic jerks [37], as the events detected in 2007.5 [38] and 2014.5 [39].

This study, together with those by [1-2] suggests that the core flows produce a signature in both magnetic and gravity variations at decadal timescale. The physical mechanisms causing the reported common variability, result in different geographical patterns at different times in magnetic and in gravity signals. The applied analysis has allowed us to retrieve the dominant time structure of the variability for each field.

Additional computations are required, in order to further investigate these correlations. Because there is no reason why the magnetic and gravity signatures should be localized at the same position on Earth, these signatures could be unsynchronized. Some methods have been proposed to study unsynchronized correlations and causalities, such as Singular Spectrum Analysis [40, and references therein], and they can be applied for such investigations. The structure of the common time variability is also worth investigating, as it can bring insight on the physical process. Investigating the new available magnetic and gravity data and developing new techniques in signal separation might allow a better localization of the phenomena, resulting in a better description of the common variability.

## 4. Discussion

The promising results on this topic lead the authors to develop the GRACEFUL project, recently selected by the ERC in the context of its SYNERGY program. The project aims at building on the well-known correlation between the magnetic field and LoD variations in addition to the above-mentioned correlation between magnetic and gravity fields. We will refine the latter correlation by improving the GRACE data analysis using all available data sets for correcting the near surface signals and estimating associated uncertainties. We will also use the most up-to-date LoD data at decadal timescale to infer the flow patterns inside the core. The core flow generated by the Earth's rotation itself will be studied in conjunction with the flow induced by the magnetic field. The effect of the CMB topography and the interactions between the core and the mantle at the CMB, and of an inner core will be taken into account to quantify induced gravity changes. Furthermore, we will use the recent core modelling tools developed in the frame of the RotaNut (Rotation and Nutation of a wobbly Earth) ERC to infer the core flows that explain all these observations. The project aims to critically improve our understanding of the liquid core dynamics from cutting-edge LOD, magnetic field, and gravity observations, as well as most comprehensive flow models capitalizing on the models and expertise developed in the frame of the ERC Advanced Grant RotaNut led in the recent years by one of us.

**Author Contributions:** Conceptualization, M.M., V.D., A.C.; investigation, M.M., V.D., A.C.; writing—review and editing, M.M., V.D., A.C. All authors have read and agreed to the published version of the manuscript.

**Funding:** The research leading to these results has received funding from the European Research Council (ERC) Advanced Grant No 670874, RotaNut – Rotation and Nutation of a wobbly Earth, as well as from the ERC GRACEFUL Synergy Grant No 855677.

**Acknowledgments:** We would like to thank Olivier de Viron for discussions and inputs in data analysis, Alejandro Blazquez for the GRACE data analysis, and Ingo Wardinski for magnetic data analysis. We also thank the three referees for very constructive suggestions to improve the manuscript.